# SVOM Gamma Ray Monitor

Dong YongWei[†], Wu BoBing, Li YanGuo, Zhang YongJie, Zhang ShuangNan

Key Laboratory of Particle Astrophysics, Institute of High Energy Physics, Chinese Academy of Sciences, Beijing 100049, China

**The Space-based multi-band astronomical Variable Object Monitor (SVOM) mission is dedicated to the detection, localization and broad-band study of Gamma-Ray Bursts (GRBs) and other high-energy transient phenomena. The Gamma Ray Monitor (GRM) onboard is designed to observe the GRBs up to 5 MeV. With this instrument one of the key GRB parameter, $E_{peak}$, can be easily measured in the hard x-ray band. It can achieve a detection rate of 100 GRBs per year which ensures the scientific output of SVOM.**

Gamma-ray Burst, $E_{peak}$, Triple phoswich

Received May 20, 2009; accepted
doi:
[†]Corresponding author (email:) dongyw@ihep.ac.cn
Supported by National Basic Research Program of China - 973 Program 2009CB824800

GRBs remain one of the most fascinating phenomena in the high energy astronomy. Several space missions, like CGRO, BeppoSAX, HETE-2 and SWIFT, have shown that GRBs are cosmological, that they are followed by long-lasting afterglows, and that they are associated with core-collapse supernovae [1]. But GRBs are far to be totally understood and solved after 40 years' research. Future GRB studies must rely on the availability of a continuous flow of accurate GRB positions (to take advantage of instrumental progress), but also on the measure of many additional parameters of GRBs (e.g. redshift, $E_{peak}$ ...) which are crucial for the understanding of the GRBs themselves and for their use as astrophysical tools.

The SVOM mission is a wide band observatory designed for making observations of GRBs from the visible band to gamma ray band. It is planned to be launched in 2013 in a circular orbit with an inclination of 30° and height of 625 km. Four instruments onboard and 3 instruments on the ground are involved. The instrumentation in space includes ECLAIRs, a hard X-ray imager and spectrometer [2], two GRM units, gamma-ray spectrometers, XIAO, a low-energy X-ray telescope, and VT, an optical telescope (See fig. 1). The ground instruments include two 1m robotic telescopes named GFT (Ground Follow-up Telescopes), and GWAC, an array of wide angle optical cameras.

The near anti-solar pointing of SVOM will allow follow-up ground instruments to measure the redshifts of the majority of SVOM GRBs.

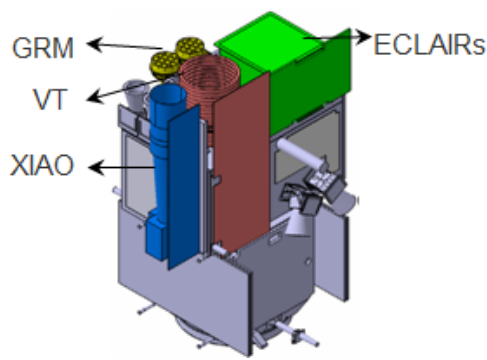

**Figure 1** the SVOM overall architecture. The Field-Of-View of ECLAIRs working in the 4–250 keV energy range is ~2 sr; that of VT is 21×21 arcmin; XIAO has a FOV of 25×25 arcmin and a focal length of 1m.

The GRM is able to provide the spectral observation on GRBs from ~30 keV to ~5000 keV. One requirement on GRM is to provide on-board GRB triggers to other payloads especially for short and hard GRBs.

Observations from few keV to few MeV are mandatory to determine as accurately as possible the GRB peak energy, a key parameter to e.g. make use of GRBs as "standard candles" and extend the measurement of the cosmological parameters at redshifts larger than 6, against $z \leq 1.7$ for type Ia supernovae. So SVOM will provide a large number of GRBs which can be standard-ized because, unlike SWIFT, it will measure the peak energy of nearly all the localized GRBs. This is also one important goal of GRM and ECLAIRs.

## 1 Instrument description and design

GRM consists of 2 identical Gamma Ray Detector (GRD) units, which are sensitive in the energy range between 30 keV and 5 MeV, are responsible for the overlap measurement with ECLAIRs instrument. Each GRD is equipped with one auxiliary calibration detector employing a plastic scintillator and an embedded $^{241}$Am source.

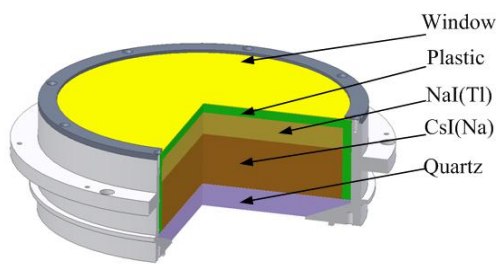

**Figure 2 A cut view of GRD scintillator case.**

The GRD detector is made up of three cylindrical scintillators creatively (See fig. 2). The NaI(Tl) crystal having a diameter of 190 mm and a thickness of 15 mm works as the main detecting element. The 35mm thick CsI(Na) crystal locates behind NaI(Tl) acting as the anti-coincidence element against photons from behind. The top and lateral sides of the two crystals are covered by a barrel-shaped organic scintillator BC448 in order to distinguish low energy charged particles from x-rays. The entrance window of the triple phoswich is made of a 2 mm thick Beryllium plate.

In order to increase the light collection capability of BC448, the three scintillators are optically glued together and viewed with one light guide who is coupled to a 5 inch PMT. The plastic scintillator is covered by specular reflector to further improve the collection efficiency. The light guide is chamfered to match the size of PMT. The 6 mm thick BC448 has enough stopping power for charged particles and the fastest pulse width thanks to its short decay time. Different pulse shape generated by scintillators can be distinguished by the Pulse Shaping Discrimination circuit (PSD). Thus the GRD has the lowest background level after removing backside photons and low energy charged particles.

A collimator made of tantalum is located in front of the scintillator case in order to suppress the background effectively. The collimator, which is separated into 16 grids by use of 1 mm thick Tantalum schemes, forms a FOV of 2 sr.

## 2 Expected outputs

The main tasks of GRM are GRB alert and $E_{peak}$ measurement each requiring high sensitivity of the instrument. The efficiency of GRM has been simulated (See fig. 3). With the cooperation of NaI and CsI, the overall efficiency reaches up to 56% at 5 MeV.

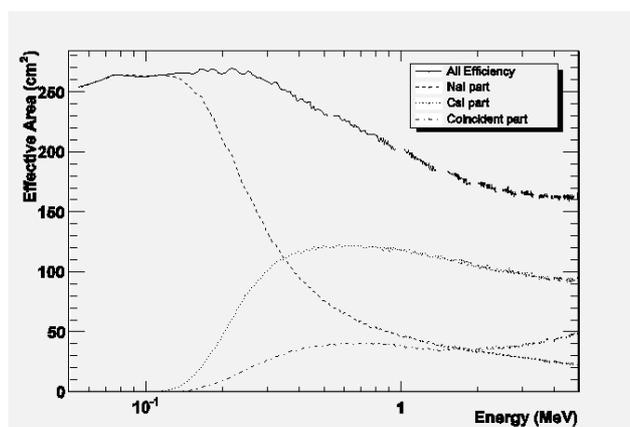

**Figure 3 Effective area of one GRD vs. the photon energy. The x-rays are emitted perpendicular to the detector plane. NaI(Tl) part represents the events only depositing energy in NaI(Tl). Coincident part corresponds to the events depositing energy in both NaI(Tl) and CsI(Tl).**

We have estimated the GRD's in-orbit background level based on the background data of BATSE/LAD. The Cosmic Diffuse Gamma Ray Background has been substituted and the other background components are scaled to the smaller size of GRM NaI(Tl) detectors. In the energy range (50,300) keV, the background of GRM is 0.61 cm$^{-2}$s$^{-1}$. The full range background is 1.36 cm$^{-2}$s$^{-1}$.

Actually the prompt hadronic component makes great contribute in the overall background especially at high energy band [3]. If we make use of the BC448, all protons <25 MeV and electrons <1.3 MeV are stopped in the plastic scintillator based on CSDA range estimation; the higher energy charged particles depositing an energy of 1 MeV in BC448 and <5 MeV in NaI(Tl) will be rejected with the help of PSD circuit; the left events will generate large signals and contribute to the dead observation time. Theoretically the entire prompt hadronic component can be eliminated from the background.

The significance of GRB is defined as:

$$\sigma(\Delta t, \Delta E) = S(\Delta t, \Delta E)/\sqrt{B(\Delta t, \Delta E)}$$

here the range of $\Delta E$ is from 0.05 MeV to 0.3 MeV, $\Delta t$ is equal to 1 s, and the GRBs detection criteria is the significance greater than 5.5 σ. Based on calculations, when $E_{peak}$ = 300 keV and the incident angle is 0°, the GRB flux threshold of two GRD units in the 1-1000 keV band is 0.90 $cm^{-2}s^{-1}$. Fig. 4 shows the sensitivity comparison between GRM and other gamma ray detectors in flight.

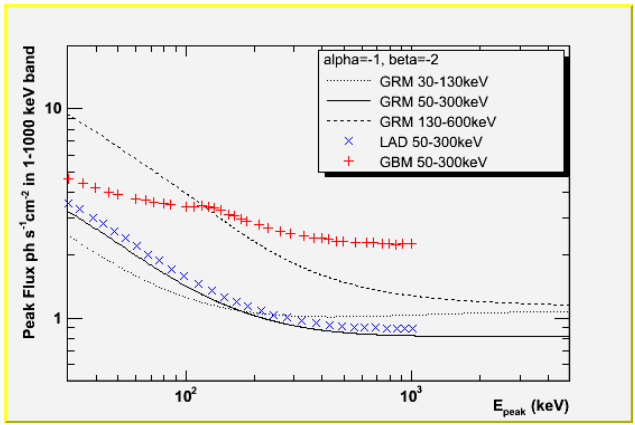

**Figure 4** Peak flux (1-1000 keV) threshold of different detectors when incident angle is 0°. The threshold significance is $\sigma_0$=5.5 except for LAD ($\sigma_0$=6.74). The parameters of BAND model used are α=-1, β=-2.

Taking the full detection efficiency and the continuous background estimation into account, The GRM sensitivity as a function of the exposure time and signal-to-noise ratio (SNR) is calculated (See Fig. 5). GRM is still sensitive to an average GRB up to 1MeV.

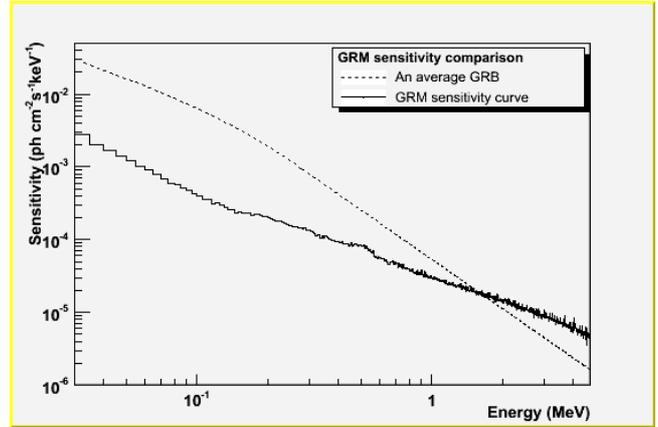

**Figure 5** GRM sensitivity at 5.5 σ for an exposure of 20 s. An average GRB means a BAND model GRB with α=-1, β=-2.25, $E_{peak}$=200 keV, a flux of 1$cm^{-2}s^{-1}$ in the 50-300 keV band.

The standard setting of GRM trigger is an energy interval of (50, 300) keV, an integration time of 1 second, and a threshold of 5.5 sigma. That result in a minimum detectable flux of 0.23 $cm^{-2}s^{-1}$ with detection efficiency taken into account. According to the equation of GRB detection capability given by Band [4]:

$$N_{sky}(\phi \geq \phi_0) = (\phi_0/0.3)^{-0.8} * 550$$

The estimated GRB detection rate of 2 GRDs will be 100/yr which meets the mission requirement.

$E_{peak}$ is the key parameter of GRB in BAND model. The average parameters of the Band spectrum are α= -1.1, β=-2.3, $E_{peak}$=266 keV [5]. Simulations of GRM data were made assuming the BATSE parameter values of GRB971110 (Fig. 6). XSPEC software is used to fit the spectrum by BAND model. An excellent agreement between the input values and the derived $E_{peak}$ result (299.8±16 keV, 1sigma) is achieved.

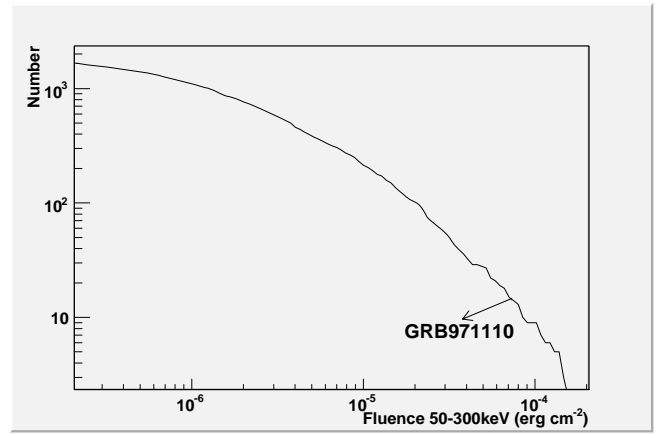

**Figure 6** LogN – log(fluence) relation from BATSE catalog with **GRB971110 marked. The BAND parameters of it are A=0.0146, α=-0.82, β=-2.21, $E_{peak}$=298 keV. The duration is 225s.**

If we take a GRB 7 times dimmer than GRB971110, the uncertainty rise up to ~60 keV with the GRD data. In comparison, a collaborated GRM-ECLAIRs fit makes the uncertainty smaller than 50 keV. The expected number brighter than such GRB is ~30 every year.

## 5 Conclusions and outlook

The joint observation of four scientific instrument onboard SVOM will help improve our understanding of the mechanism of GRBs and the cosmology physics. The GRM extending the energy range up to several MeV greatly increases the possible GRB output. The optimization design, including the collimator and the triple phoswich with particle-shielding function, reduces the background of GRM to the lowest level. Also, the $E_{peak}$ parameter is well determined by fitting a wide-band spectrum especially when the ECLAIRs data are included. A major improvement on the GRB science is foreseen with the successful operation of SVOM in 2010s.